\documentstyle[preprint,aps]{revtex}

\begin{document}
\draft
\title{Gravitational Thermodynamics of Space-time Foam in One-loop Approximation}
\author{Liao Liu\thanks{%
E-mail address: liuliao@class1.bao.ac.cn} and Yongge Ma\thanks{%
E-mail address: ygma@263.net}}
\address{Department of Physics and Institute of Theoretical Physics,\\
Beijing Normal University, Beijing 100875, P.R. China}
\maketitle

\begin{abstract}
We show from one-loop quantum gravity and statistical thermodynamics that
the thermodynamics of quantum foam in flat space-time and Schwarzschild
space-time is exactly the same as that of Hawking-Unruh radiation in thermal
equilibrium. This means we show unambiguously that Hawking-Unruh thermal
radiation should contain thermal gravitons or the contribution of quantum
space-time foam. As a by-product, we give also the quantum gravity
correction in one-loop approximation to the classical black hole
thermodynamics.
\end{abstract}

\pacs{PACS number(s): 04.60.Gw, 04.60.Dy, 05.30.Ch}

The space-time foam-like structure (FLS) was firstly proposed by J. A.
Wheeler about forty years ago[1]. He argued that the space-time may have a
multiple connected non-trivial topological structure in Planck scale, though
it seems smooth and simply connected in the large. The possible influence of
FLS on field theory and the thermodynamical properties of FLS itself have
been discussed by many authors[2-9]. In this paper, we would like to discuss
the thermodynamical properties of FLS only from one-loop quantum gravity and
statistical thermodynamics, so the result we get may be much more reliable.

If time in Euclidean quantum gravity has an imaginary period of $i\beta
=i(1/T)$, (henceforth, we take $\hbar =c=G=k=1$) then the partition function 
\begin{equation}
Z=\sum_n\exp (-\beta E_n)
\end{equation}
of canonical ensemble can be rewritten as an Euclidean path integral 
\begin{equation}
Z=\int D(g,\phi )\exp (-\hat{I}(g,\phi )),
\end{equation}
where $\hat{I}(g,\phi )$ is the Euclideanized action of gravity, $g$, and
matter field, $\phi $, and $E_n$ is the $n$-th energy eigenvalue of certain
differential field operator on its eigenstate vector $\left| g,\phi
\right\rangle _n$. For the pure quantum gravity case, we put $\phi =0$ and 
\begin{equation}
g_{ab}=g_{ab}^{(0)}+\bar{g}_{ab},
\end{equation}
where $\bar{g}_{ab}$ is the metric fluctuation of the background metric $%
g_{ab}^{(0)}$, then we can expand the action in Taylor series about the
background field $g^{(0)}$ as 
\begin{equation}
\hat{I}(g)=\hat{I}(g^{(0)})+\hat{I}_2(\bar{g})+[\text{higher-order terms}],
\end{equation}
where $\hat{I}_2(\bar{g})$ is the well known one-loop term of the Euclidean
gravitational action. In one-loop approximation, the logarithm of partition
function, $Z$, reads 
\begin{equation}
\ln Z=-\hat{I}(g^{(0)})+\ln \int D(\bar{g})\exp (-\hat{I}_2(\bar{g})).
\end{equation}
As $\hat{I}(g^{(0)})$ is equal to the Gibbons-Hawking's surface term for
vacuum Einstein gravity without cosmological term, the contributions to $\ln
Z$ come from the surface term and $\hat{I}_2(\bar{g})$ in one-loop
approximation.

Now from quantum gravity and statistical thermodynamics we try to study the
gravitational thermodynamics of the one-loop quantum gravity for flat
space-time and Schwarzschild space-time background. Let us consider the
gravitational field inside volume $V$ and with an imaginary time period $%
i\beta $. Hawking showed exactly that $\ln Z$ in one-loop approximation with
flat space-time background reads[10] 
\begin{equation}
\ln Z=\frac{4\pi ^3r_0^3T^3}{135}=\frac{\pi ^2}{45}\beta ^{-3}V
\end{equation}
for a system at temperature $T=\beta ^{-1}$, contained in a spherical box of
radius $r_0$, where the Casimir effect of the finite size of volume $V$ is
neglected. (Note that the factor $4\pi ^5$ in Eq.(15.99) of Hawking's
original paper [10] should be corrected as $4\pi ^3$ in Eq.(6). ) Hawking
argued that Eq.(6) is just the contribution of the thermal gravitons to the
partition function. However, in our opinion, if FLS is created from metric
fluctuation, an equivalent interpretation of Eq.(6) as the contribution of
FLS can also be given. Let $P_n$ be the probability of FLS in volume $V$ in
the $n$-th energy eigenstate, then from 
\begin{equation}
S=-\sum_nP_n\ln P_n
\end{equation}
and 
\begin{equation}
P_n=Z^{-1}\exp (-\beta E_n),
\end{equation}
the entropy of FLS in $V$ is given by 
\begin{equation}
S=\beta <E>+\ln Z,
\end{equation}
where the expected value of energy is given by 
\begin{equation}
<E>=-\frac \partial {\partial \beta }\ln Z.
\end{equation}
From Eqs. (6), (9) and (10) it is easy to show that the entropy, $S$, and
energy, $U$, of FLS inside volume $V$ are respectively 
\begin{equation}
S=\frac{4\pi ^2}{45}\beta ^{-3}V
\end{equation}
and 
\begin{equation}
U=<E>=\frac{\pi ^2}{15}\beta ^{-4}V.
\end{equation}
So, the entropy density, $\rho _S$, and energy density, $\rho _U$, of FLS
are respectively 
\begin{equation}
\rho _S=\frac{4\pi ^2}{45}T^3
\end{equation}
and 
\begin{equation}
\rho _U=\frac{\pi ^2}{15}T^4.
\end{equation}
They are exactly the same as that of the black body radiation. As is known,
the time coordinate of inertial system in flat space-time has no imaginary
period, or the same, the period $\beta $ in inertial system is infinite.
Hence the temperature of FLS is always zero for a inertial observer.
However, the only case that the time coordinate in flat space-time has a
finite imaginary period is that of Rindler system. This is clear from the
coordinate transformation from inertial coordinates ($t,x,\theta ,\varphi $)
to Rindler coordinates ($\eta ,\xi ,\theta ,\varphi $), 
\begin{equation}
\begin{array}{c}
t=a^{-1}e^{a\xi }sh(a\eta ) \\ 
x=a^{-1}e^{a\xi }ch(a\eta )
\end{array}
.
\end{equation}
Evidently, $\eta $ has an imaginary period of $2\pi /a$ after it is
Euclideanized, i.e., $\eta \rightarrow -i\eta .$ So, $\beta =2\pi /a$ and
the famous Unruh temperature[11,12] 
\begin{equation}
T_U=\beta ^{-1}=\frac a{2\pi }
\end{equation}
results.

The above discussion suggests that, though the temperature of FLS for
inertial observer is zero, but the Rindler observer will find himself
immersed in a heat bath of temperature $T_U$ of FLS. Now a problem confront
us is whether we are really in an inertial system or in a Rindler system
with heat bath? It seems no choice can be given a priori. The only
reasonable choice depends on whether we can measure out the temperature of
FLS or not. As is known, there is an universal background black body
radiation of $\symbol{126}3^0k$ everywhere in the universe. If the idea of
thermal gravitons or thermal FLS is not wrong, the densities $\rho _S$ and $%
\rho _U$ may be too small compared with that of the $\symbol{126}3^0k$
background radiation, so that a measurement of them can hardly be given,
especially, if we remember that in order to get an Unruh temperature of $%
1^0k $ the proper acceleration of the Rindler observer should approximately
be[13] 
\begin{equation}
\alpha ^{-1}=ae^{-a\xi }\cong 2.4\times 10^{20}m/s^2\cong 10^{19}g_E,
\end{equation}
where $g_E$ is the proper acceleration on the surface of the earth. Hence,
it seems highly impossible that FLS can have any measurable thermal
properties in practice.

Hawking also showed that the one- loop approximation of the logarithm of the
partition function, $\ln Z,$ for pure gravity in Schwarzschild black hole
background metric is given approximately by[10] 
\begin{equation}
\ln Z=-\hat{I}(g^{(0)})+\ln \int D(\bar{g})\exp (-\hat{I}_2(\bar{g}))=-\frac{%
\beta ^2}{16\pi }+\frac{106}{45}\ln (\frac \beta {\beta _0})+\frac{4\pi
^3r_0^3}{135\beta ^3}+O(r_0^2\beta ^{-2}),
\end{equation}
where $\hat{I}(g^{(0)})=\beta ^2/16\pi $ is the Gibbons-Hawking's surface
term of Schwarzschild space-time, $\beta =T^{-1}=8\pi M$ is just the
imaginary time period of Schwarzschild black hole, $\beta _0$ is an
arbitrary constant of energy dimensionality, and $r_0$ is the proper radius
of a spherical box enclosing the Euclidean section of a Schwarzschild black
hole in its center. From Eqs. (9) (10) and (18), we can get the
gravitational entropy, $S,$ and energy, $U,$ of the whole system of proper
volume $V$ as 
\begin{equation}
U=-\frac \partial {\partial \beta }\ln Z=M-\frac{53}{180\pi M}+\frac{\pi ^2}{%
15}T^4V,
\end{equation}
\begin{equation}
S=\beta <E>+\ln Z=4\pi M^2+\frac{106}{45}[\ln (\frac M{M_0})-1]+\frac{4\pi ^2%
}{45}T^3V,
\end{equation}
where $M_0\equiv \beta _0/8\pi $.

The last terms in the right-hand sides of Eqs. (19) and (20) are exactly the
same as that of the Hawking radiation in equilibrium, that can only be
ascribed to the contribution of thermal gravitons or of quantum foam, while
the first terms are the familiar contribution of classical Schwarzschild
black hole, and the second terms which originate from one-loop quantum
gravity are no doubt the quantum gravity correction to the classical
Schwarzschild black hole. It is interesting to note that, the quantum
gravity correction of black hole entropy is always negative and inverse
proportional to the black hole mass, while the quantum gravity correction of
black hole entropy is negative, positive or zero in the case $\frac M{M_0}<e$%
, $>e$, or $=e$.

In summary, the gravitational thermodynamics of FLS in one-loop
approximation is exactly the same as that of Hawking-Unruh radiation in
thermal equilibrium. In other words, we show unambiguously that the
Hawking-Unruh thermal radiation not only contain the matter particles but
also contain the thermal gravitons or the contribution from the quantum FLS
of space-time. We show also that the one-loop quantum gravity gives quantum
correction to the classical thermodynamics of Schwarzschild black hole.

\acknowledgments 

This work is supported by the National Natural Science Foundation of China
under Grand No. 19473005.

\end{document}